\documentclass[journal,onecolumn]{IEEEtran}

\usepackage{graphicx}
\usepackage[table,xcdraw]{xcolor}
\usepackage{framed}
\usepackage{enumitem}

\definecolor{formalshade}{rgb}{0.95,0.95,1}
\definecolor{darkblue}{rgb}{0.145, 0.118, 0.580}

\newenvironment{formal}{%
  \MakeFramed{\advance\hsize-\width\FrameRestore}%
  \noindent\hspace{-4.55pt}

  \vspace{2pt}\vspace{2pt}%
}
{%
  \vspace{2pt}\endMakeFramed%
}

\begin{document}

\title{Personalized Education with Generative AI and Digital Twins: VR, RAG, and Zero-Shot Sentiment Analysis for Industry 4.0 Workforce Development}

\author{\IEEEauthorblockN{Yu-Zheng Lin\IEEEauthorrefmark{1}, 
Karan Petal\IEEEauthorrefmark{2},
Ahmed H Alhamadah\IEEEauthorrefmark{2}, Sujan Ghimire\IEEEauthorrefmark{2}, Matthew William Redondo\IEEEauthorrefmark{2}, \\
David Rafael Vidal Corona\IEEEauthorrefmark{3}, Jesus Pacheco\IEEEauthorrefmark{3}, Soheil Salehi\IEEEauthorrefmark{1}, and Pratik Satam\IEEEauthorrefmark{2}}\\
\IEEEauthorblockA{\IEEEauthorrefmark{1}Department of Electrical and Computer Engineering, University of Arizona, Tucson, AZ, USA\\
\IEEEauthorrefmark{2}Department of Systems and Industrial Engineering, University of Arizona, Tucson, AZ, USA\\
\IEEEauthorrefmark{3}Department of Industrial Engineering, University of Sonora, Hermosillo, Mexico\\
Email: \{\IEEEauthorrefmark{1}yuzhenglin, \IEEEauthorrefmark{2}karanpatel, 
\IEEEauthorrefmark{2}alhamadah,
\IEEEauthorrefmark{2}sghimire,
\IEEEauthorrefmark{2}mredondo245,
\IEEEauthorrefmark{1}ssalehi, 
\IEEEauthorrefmark{2}pratiksatam\}@arizona.edu; 
\{\IEEEauthorrefmark{3}david.vidal, \IEEEauthorrefmark{3}jesus.pacheco\}@unison.mx}
}

\markboth{This paper accepted by ASEE 2025 Annual Conference}%
{}

\maketitle

\begin{abstract}
While the advent of the Fourth Industrial Revolution (4IR) technologies, like cloud computing, machine learning, and artificial intelligence have brought convenience and productivity improvements, they have also introduced new challenges in training and education that require the reskilling of existing employees and the building of a new workforce. Exacerbated by the already existing workforce shortages, this mammoth workforce reskilling and building effort aims to build a high-tech workforce capable of operating and maintaining these 4IR systems; requiring a higher student retention and persistence. This increase in student retention and persistence will be especially critical when training the workforce originating from marginalized communities like Underrepresented Minorities (URM), where challenges arise due to lack of access to high-quality education throughout the trainee’s formative years (pre/middle/high schools), creating a cyclic set of knowledge dependencies that are difficult to meet.
To address these challenges, this research presents Generative AI-based Personalized Tutor for Industrial 4.0 (gAI-PT4I4), a framework that focuses on personalization of 4IR experiential learning, using sentiment analysis to gauge student's knowledge comprehension, while using a combination of generative AI and finite automaton to personalize the content to the students' learning needs. The framework administers experiential learning, using low-fidelity Digital Twins that enable virtual reality-based (VR) training exercises focusing on 4IR training. The VR environment, integrates a generative AI teaching assistant called the Interactive Tutor, that guides the student through the training exercises, with audio and text communications. The gAI-PT4I4 uses these audio/text communications between the tutor and the trainee to perform sentiment analysis using a novel zero-shot learning pipeline, that uses Large Language Models (LLMs), with prompt engineering to evaluate student sentiments in teacher-student conversations without any prior training (unlike traditional deep learning techniques). Our experimental evaluation shows this zero shot sentiment analysis pipeline built on GPT4 has an 86\% accuracy while classifying student-teacher interactions as positive or negative. The gAI-PT4I4, explores the use of retrieval-augmented generation (RAG), for grounded generation of personalized learning content using domain specific knowledge in combination with LLMs to provide personalized teaching in real-time.  Lastly the gAI-PT4I4, uses finite automaton to split each exercise in different states of varying difficulty, requiring 80\% student task-performance accuracy, to dynamically increase the difficulty of the exercise by transitioning through the automaton. Experimental evaluation with a group of 22 volunteers showed the participant skills improvement from an accuracy to over 80\%, with reduction in training time. Lastly, this paper also presents a Multi-Fidelity Digital Twin model, that presents a scalable framework to map Digital Twin functionalities with different education levels, mapping Bloom's Taxonomy, and the Kirkpatrick's model with levels of Digitial Twin complexity.   
\end{abstract}


\IEEEpeerreviewmaketitle

\section{Introduction}
\IEEEPARstart{T}{he} rapid adoption of automation in cyber-physical systems (CPS), such as smart manufacturing plants, is transforming the world and driving the Fourth Industrial Revolution (4IR). These 4IR systems rely on critical technologies like cloud computing, machine learning (ML), artificial intelligence (AI), and universal network connectivity to achieve increased productivity and sustainable growth with agile and responsive supply chains \cite{briggs2019beyond, schwab2016fourth, bhamare2020cybersecurity}. While these technologies bring significant benefits, their growing adoption has also increased the complexity of manufacturing systems, making them increasingly difficult to manage, secure, and optimize. These transformative changes make it critical for the 4IR workforce to have a strong understanding of topics in 4IR, requiring reskilling of the existing workforce in addition to training a new workforce, a mammoth task on scale \cite{Manufacturing_Skill}. However, such engineering training programs face multiple obstacles. For instance, although online training programs are cost-effective, easy to scale, and are preferred for reskilling/upskilling efforts \cite{ginder2019enrollment}, 4IR workforce training requires access to specialized hardware, making such programs unsuitable for online offerings. Furthermore, the current labor shortages \cite{us_senate_2017} will require new workforce creation programs or reskilling/upskilling programs to have higher student retention and persistence while meeting the programs' learning objectives, a challenge difficult to overcome. Researchers have shown that student background and upbringing play a role in student persistence through STEM programs, wherein data shows that it is especially challenging for students from marginalized communities like underrepresented minorities (URMs) to succeed. These difficulties can be attributed to a lack of access to high-quality education throughout the student’s formative years (pre/middle/high schools), creating a cyclic set of knowledge dependencies that are difficult to meet \cite{ogbu2022voluntary, wladis2015stem, shih2024challenges}.

To address these challenges, this study will explore the innovative use of generative AI (gAI) with VR and sentiment analysis, aiming to build a personalized learning environment for the training of the 4IR workforce \cite{osunbunmi2024board, lee2012augmented}. Such personalized courses will adapt to a student’s learning style and background, ensuring skill building meets the program’s learning objectives. Furthermore, this study will explore the use of gAI-based interactive instructors to increase an engineering student's sense of belonging, engineering identity, and academic attainment and improve student retention and course completion. This study presents the gAI-based Personalized Tutor for Industrial 4.0 (gAI-PT4I4), as shown in Figure \ref{fig:gAI-PT4I4}, an interactive framework to provide gamified virtual reality (VR) based cybersecurity training that personalizes the coursework through sentiment analysis and generative AI, to ensure meeting of the course’s learning outcomes while improving student retention through bolstering their engineering identity and sense of belonging.

The main contributions of this paper are as follows:
\begin{itemize}
    \item \textbf{Multi-Fidelity Digital Twin Education Reference Model:} We propose a multi-fidelity digital twin and generative AI reference model for education. This model defines a multi-fidelity digital twin and combined it with Bloom's Taxonomy and Kirkpatrick model to provide a reference for digital twin applications on different education needs and learning performance evaluation indicators for the use of generative AI.
    \item \textbf{Develop an immersive 4IR learning interface using low-fidelity Digital Twins:} We designed and implemented a learning platform, integrating low-fidelity digital twin technology and LLMs to provide an immersive learning experience, including smart manufacturing factory tours, personal protective equipment inspection training, etc. 
    \item \textbf{Zero-shot gAI-based sentiment analysis:} This study uses prompt engineering to implement zero-shot sentiment analysis technology to classify conversation sentiment and convert qualitative sentiment into quantitative data. In addition, we evaluate this framework on two separate datasets, 1) Teacher-Student conversation, and 2)Twitter Sentiment Analysis Training Corpus (TSATC) dataset collected from Twitter, highlighting the higher performance accuracy of our approach compared to tradition neural networks (NNs) without any prior training.
    \item \textbf{Create a sentiment analysis dataset for teacher-student dialogues:} We labeled more than a thousand teacher-student dialogues from the Google Education Dialogue Dataset, creating a new sentiment labeled dataset called EduTalk Sentiment Dataset, to test and verify the effectiveness of the zero-shot gAI-based sentiment analysis pipeline. Our EduTalk Sentiment Dataset can be found in Reference \cite{GEDD-S}.
     \item \textbf{Enhanced LLMs’ responses in 4IR expertise with GraphRAG as personalized tutors:} This study uses GraphRAG to improve the precision of LLM responses in 4IR expertise to provide flexible course content adjustment
    \item \textbf{Adaptive Difficulty Mechanism with Finite Automaton:} We use finite automaton in gAI-PT4I4 to split each exercise into different states of varying difficulty, requiring 80\% student task-performance accuracy, to dynamically increase the difficulty of the exercise by transitioning through the automaton.
\end{itemize}

\section{Methods}

\subsection{Multi-Fidelity Digital Twin Education Framework}
\begin{figure*}[!t]
  \centering
  \includegraphics[width=0.8\linewidth]{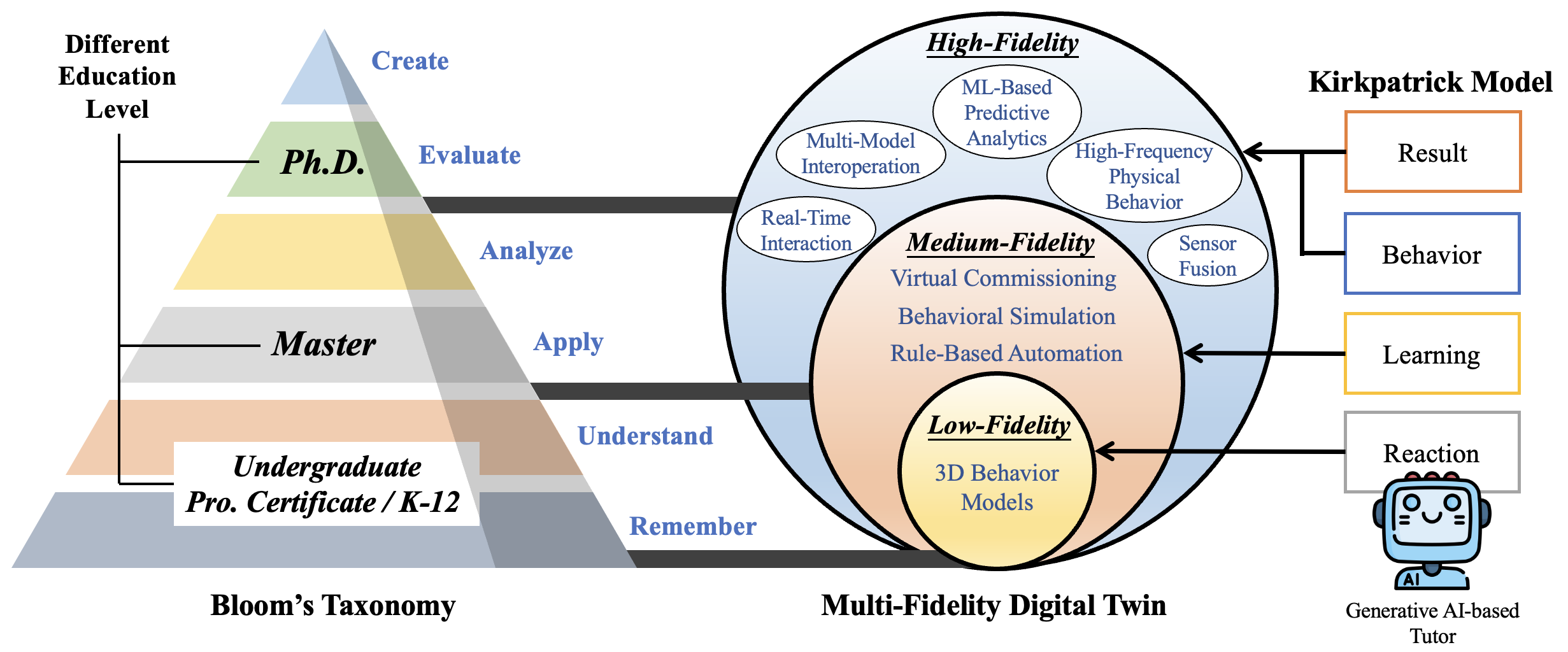}
    \caption{Multi-Fidelity Digital Twin Education Reference Model}
    \label{fig:ref_model}
\end{figure*}

Digital twins (DT) originated in the aerospace field \cite{glaessgen2012digital} and today impact a wide range of fields such as transportation, climate, manufacturing, etc \cite{kuvsic2023digital, bauer2021digital, lin2023dt4i4}, and also education \cite{razzaq2023deepclassrooms}, wherein "Software systems replicating the behavior of one or more physical processes using one or more behavior models \cite{lin2023dt4i4, lin2024transforming}." The Multi-Fidelity Digital Twin Education Framework, maps different DT functionalities to education levels, and training needs based on the Bloom's Taxonomy and the Kirckpatrick model; presenting a scalable approach to define software design requirements based on learning outcomes \cite{smidt2009kirkpatrick, krathwohl2002revision}, as illustrated in Figure \ref{fig:ref_model}. Bloom's taxonomy's six cognitive levels- Remember, Understand, Apply, Analyze, Evaluate, and Create - are assigned to different stages of education: Undergraduate, Master, and Doctoral degrees based on their learning outcomes. In the presented Multi-Fidelity Digital Twin Education framework, we used multi-fidelity digital twin as a core concept and mapping to different education levels with Bloom's Taxonomy, then used LLM based on the Kirkpatrick model to evaluate the learning performance.
Applying the Multi-Fidelity Digital Twin Framework to the 4IR workforce development challenge: 
\begin{itemize}
    \item For undergraduate students and certificate programs, Low-Fidelity Digital Twins (e.g., 3D behavior models) help establish 4IR basic knowledge to understand cyber-physical systems, enabling students to grasp key concepts such as automation, data integration, and the role of emerging technologies in modern industries. At this stage, gAI-based tutor can evaluate the ``Reaction'' based on the Kirkpatrick model, which is whether the student training is favorable, engaging, and relevant to their jobs.
    \item For master student level, Medium-Fidelity Digital Twins (e.g., virtual commissioning and behavioral simulations) facilitate the application of learned concepts and analytical skills. Students can observe changes in device behavior through parameter adjustments and analyze their impact on output. At this stage, the gAI-based tutor can evaluate ``Learning'' based on the Kirkpatrick model, which is the acquisition of the desired knowledge, skills, and confidence based on their participation in the training.
    \item For doctoral student level, High-Fidelity Digital Twins (e.g., real-time interaction, machine learning predictive analytics, and multi-model interoperability) enable students to engage in creative problem-solving and advanced evaluation. At this stage, the gAI-based tutor can evaluate ``Behavior'' and ``Result'' based on the Kirkpatrick model, which is students apply what they learned during training when they are doing research and their research results.
\end{itemize}

\subsection{Generative AI-based Personalized Tutor for Industrial 4.0 (gAI-PT4I4) Framework}
\begin{figure*}[!h]
  \centering
    \includegraphics[width=0.8\linewidth]{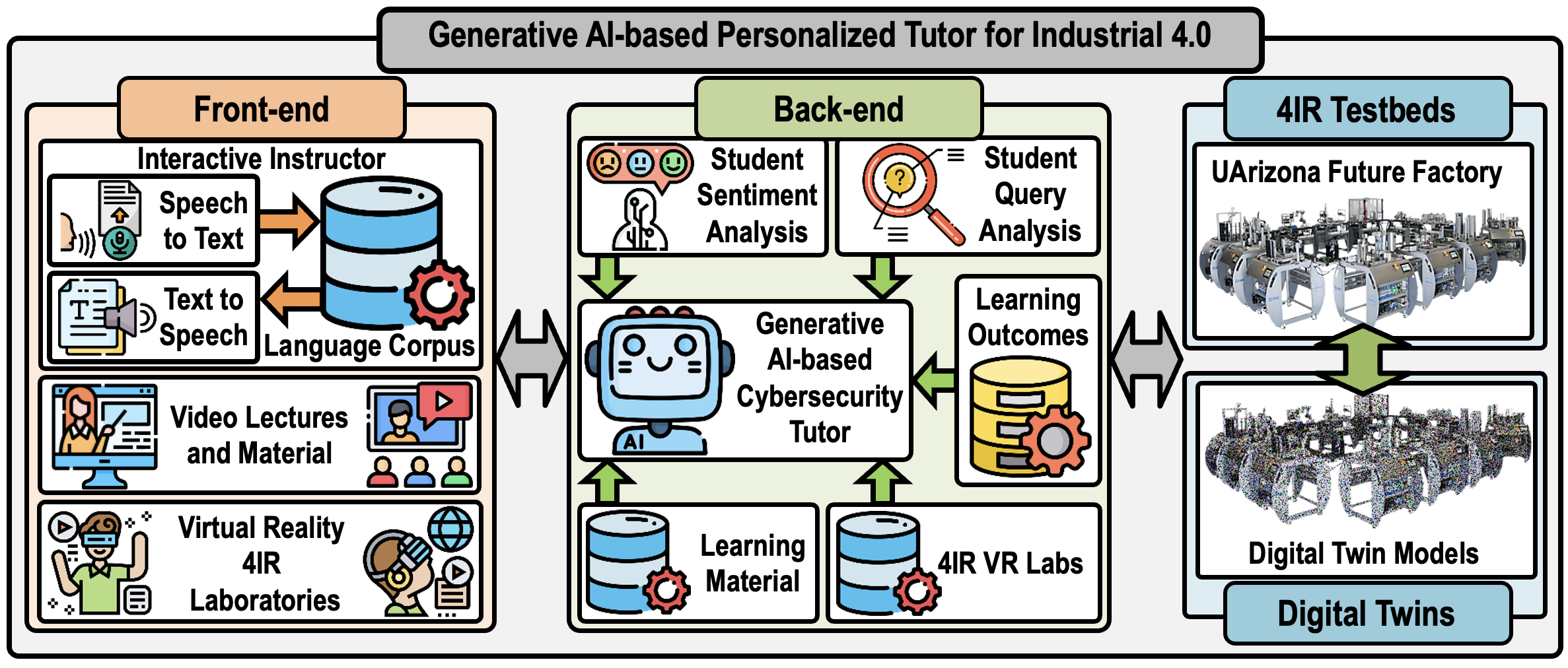}
    \caption{Generative AI-based Personalized Tutor for Industrial 4.0 (gAI-PT4I4) Framework}
    \label{fig:gAI-PT4I4}
\end{figure*}

The proposed Generative AI-based Personalized Tutor for Industrial 4.0 (gAI-PT4I4) Framework uses and implements the Multi-Fidelity Digital Twin Education Framework, for undergraduate students and professional certificate programs via low fidelity DTs for 4IR workforce development as illustrated in Figure \ref{fig:gAI-PT4I4}. 

\subsection{Virtual Reality as a Learning Interface}
This learning interface aims to provide a 4IR training environment by integrating the photogrammetry method proposed by Ahmed et al. \cite{alhamadah2024photogrammetry} and the DDD-GenDT method proposed by Lin et al. \cite{lin2024ddd} using a Virtual Reality (VR) environment with microservice architecture to provide virtual equipment and their behaviors to users. VR creates an immersive training experience with specialized hardware typically used in 4IR environments, allowing realistic experimentation, including scenarios with catastrophic failures without inherent risks to equipment or human life \cite{hamilton2021immersive, carruth2017virtual}. For example, while studying 4IR cybersecurity \cite{satam2020wids, tunc2015claas}, students can be transported to a manufacturing factory floor through the VR environment, where they can perform and observe the impacts of data injection attacks on industrial robots \cite{satam2023cps, satam2020wids, shao2021multi} facility meaningful learning without risking student on the equipment. The learning in the VR-based Industrial 4.0 Labs introduces the Interactive Instructor (I2), who will provide students with deeper insight into the lecture topics and lab activities, correct students' mistakes, and guide students into becoming better 4IR professionals. The I2 will also use its interactions with the student to personalize the lectures and lab activities for that particular student's background, learning style, learning speed, and proficiency. 

Moreover, we used a finite automaton in the gaming mechanism to dynamically adjust the difficulty levels based on the student's performance in VR experiential learning activities. For example, if a student consistently exceeds performance thresholds, the automaton transitions the system to a higher difficulty state, introducing more complex scenarios. Conversely, if the student struggles to meet the thresholds, the automaton adjusts the difficulty downward, simplifying tasks to ensure that the student remains engaged without feeling overwhelmed.

\subsection{Leveraging Generative AI-based Virtual Tutors for Personalized Learning}
Integrating a virtual tutor based on generative AI provides instant personalized guidance based on the needs of the learners. Using the design advantages of LLMs for solving generalized tasks,  the virtual tutor will continuously monitor the learner's input and the learning environment to evaluate performance, provide feedback, make suggestions for improvement, and make up for deficient knowledge. We achieve this goal in two parts. 

\begin{figure*}[!h]
  \centering
  \includegraphics[width=.9\linewidth]{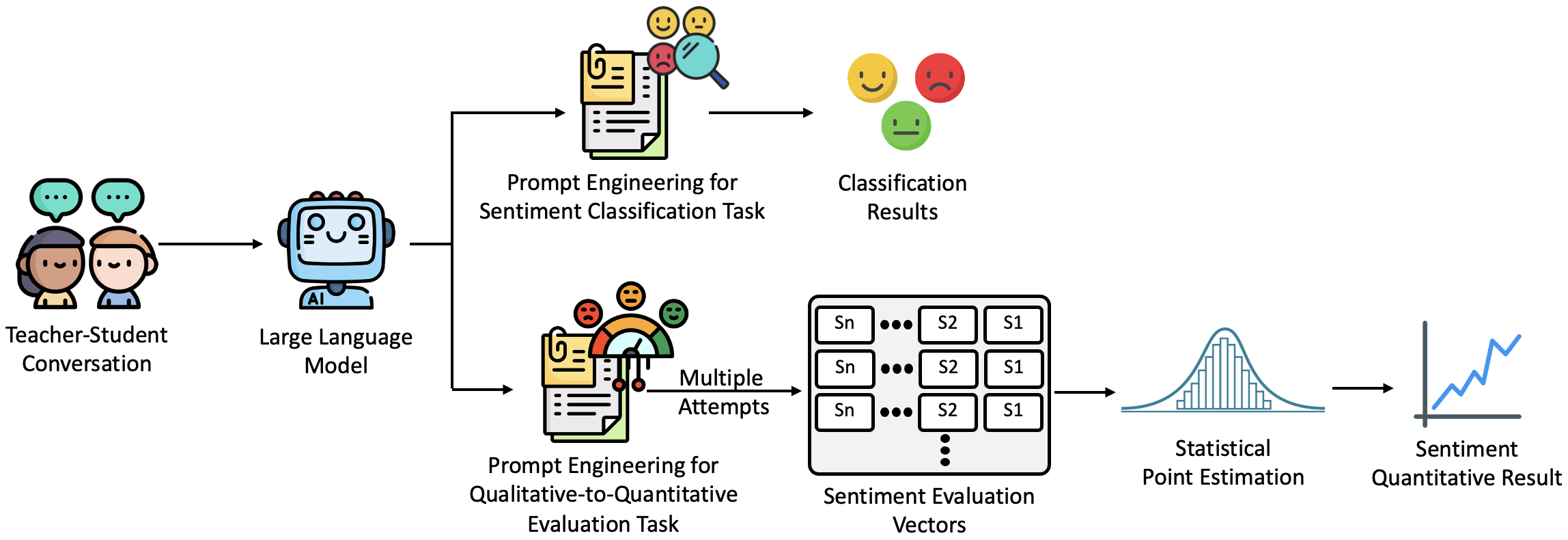}
    \caption{Zero-shot LLM-based sentiment analysis pipeline}
    \label{fig:llm_sentiment}
\end{figure*}

The first part uses zero-shot sentiment analysis, as shown in Figure \ref{fig:llm_sentiment}, which has the advantage of not relying on a large amount of labeled data in a specific field and can be quickly applied to new tasks or different scenarios, significantly reducing development and training costs. Through well-craft prompt engineering, we enable the model to understand the task context and generate reasonable results accurately \cite{white2023prompt}. We divide the task into classification tasks and qualitative-to-quantitative evaluation tasks in sentiment analysis. The classification task mainly analyzes the polarity of sentiment, such as ``positive'' or ``negative.'' The model evaluates the learner's sentiment state by examining the learner's input text and makes a comprehensive judgment based on other parameters of the learning environment. The qualitative-to-quantitative task converts the sentiment into specific quantitative indicators, such as expressing the positive and negative of the learner's sentiment in numerical form to assess the mutual influence of teacher-student interaction during the dialogue process.

The second part is the integration of LLMs with Retrieval-Augmented Generation (RAG) \cite{lewis2020retrieval} to dynamically access domain-specific knowledge, making its responses more precise and contextually relevant. RAG technology allows the model to retrieve relevant information from external databases before generating answers, thereby enhancing the model's knowledge base, especially when addressing specific technical issues. The virtual tutor can provide more accurate and particular guidance by establishing a database with specific domain knowledge, machine-operation guides, troubleshooting methods, and frequently asked questions. Such responses will be personalized, and LLMs can bridge student knowledge gaps to achieve instructional objectives based on student feedback through content and sentiment analysis.

To evaluate the LLMs' performance on sentiment analysis in an education scenario. We manually labeled more than 1,000 data from the Google Education Dialogue Dataset \cite{shani2024multi}, called EduTalk Sentiment Dataset \cite{GEDD-S}, as a refined data set for this study, where each dialogue instance is annotated with sentiment labels that capture the emotional
tone of the conversation, such as positive or negative emotions. In addition, we also considered that students may use intuitive Internet slang to interact with the instructor when taking online courses. We used the TSATC testing dataset \cite{paws2019naacl} collected from Twitter to examine the effectiveness of our LLM-based zero-shot sentiment analysis method.

\section{Results}

\subsection{Learning Interface build with Unity}
The Unity-based learning interface provides an immersive and interactive learning experience. The initial version of the interface consists of four modules that utilize low-fidelity digital twins, each focusing on a specific aspect of Industry 4.0 environments and skills development. Students can explore realistic manufacturing scenarios through these modules, interact with virtual components, and enhance their knowledge and decision-making skills in a gamified and engaging environment.

\begin{figure}[h!]
    \centering
    \begin{minipage}{0.45\textwidth}
        \centering
        \includegraphics[width=\textwidth]{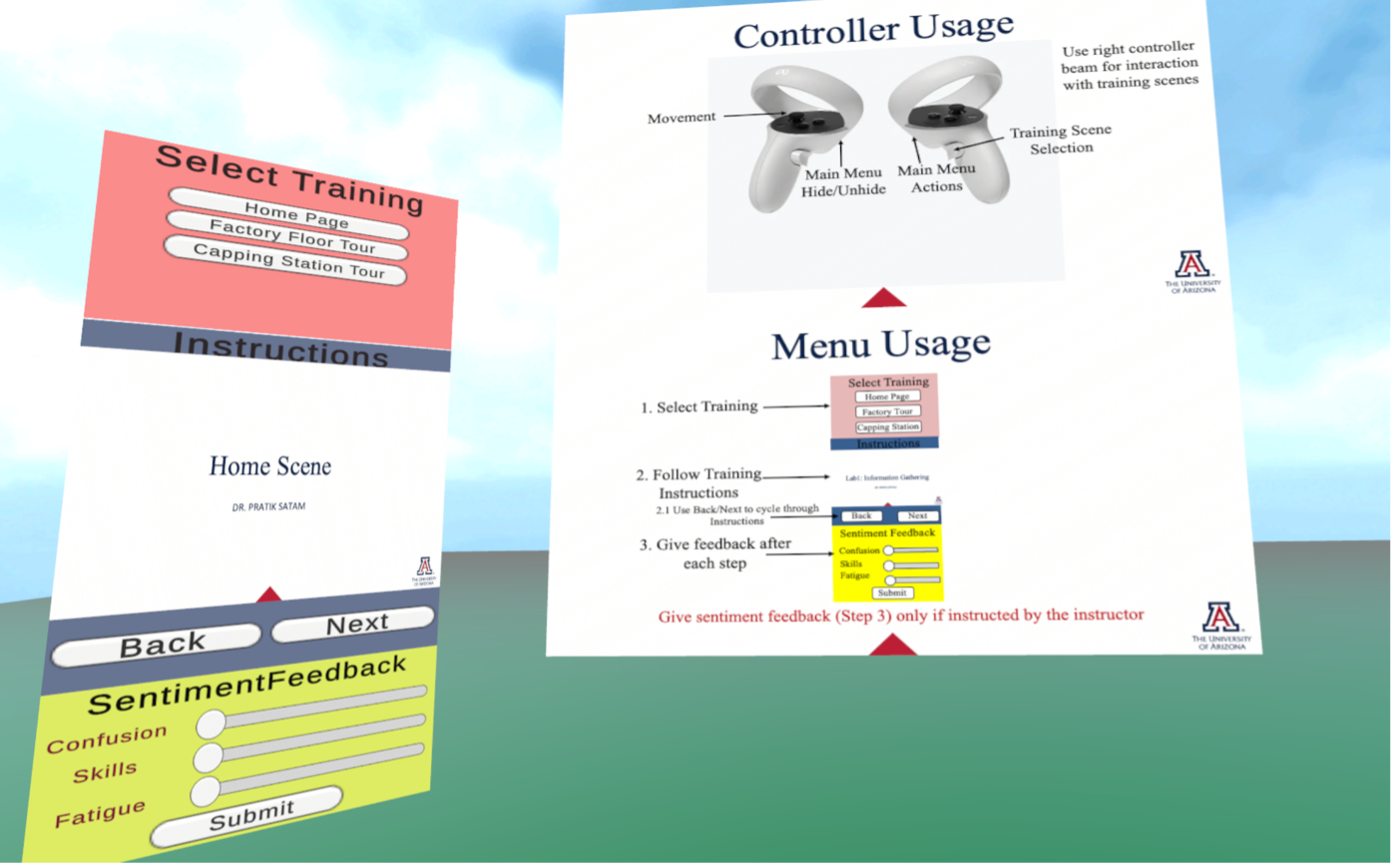}
        \makebox[0.45\textwidth]{(a) Module 1: Home Scene}
        \label{fig:unity_module1}
    \end{minipage}
    \hfill
    \begin{minipage}{0.45\textwidth}
        \centering
        \includegraphics[width=\textwidth]{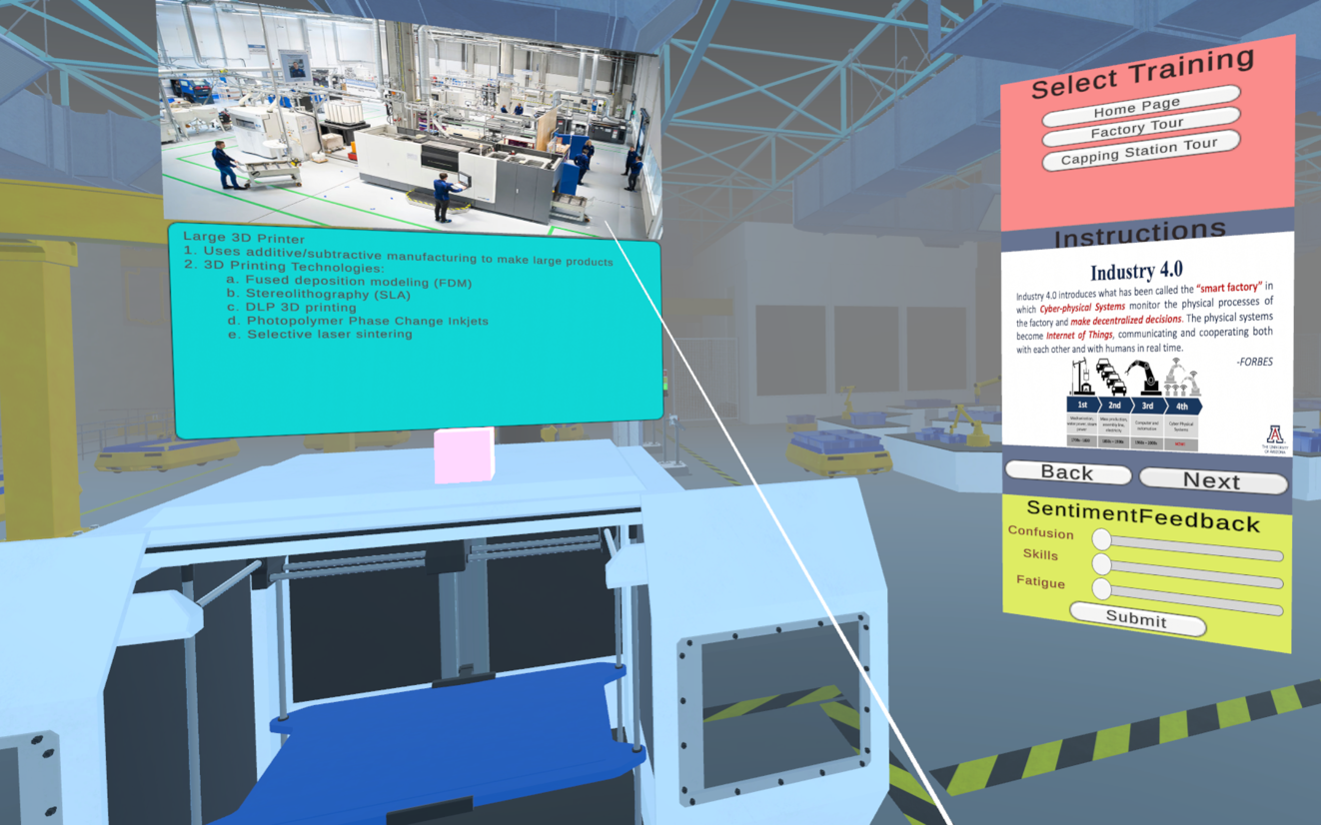}
        \makebox[0.45\textwidth]{(b) Module 2: Factory Floor Tour}
        \label{fig:unity_module2}
    \end{minipage}

    \vspace{5pt} 
    
    \begin{minipage}{0.45\textwidth}
        \centering
        \includegraphics[width=\textwidth]{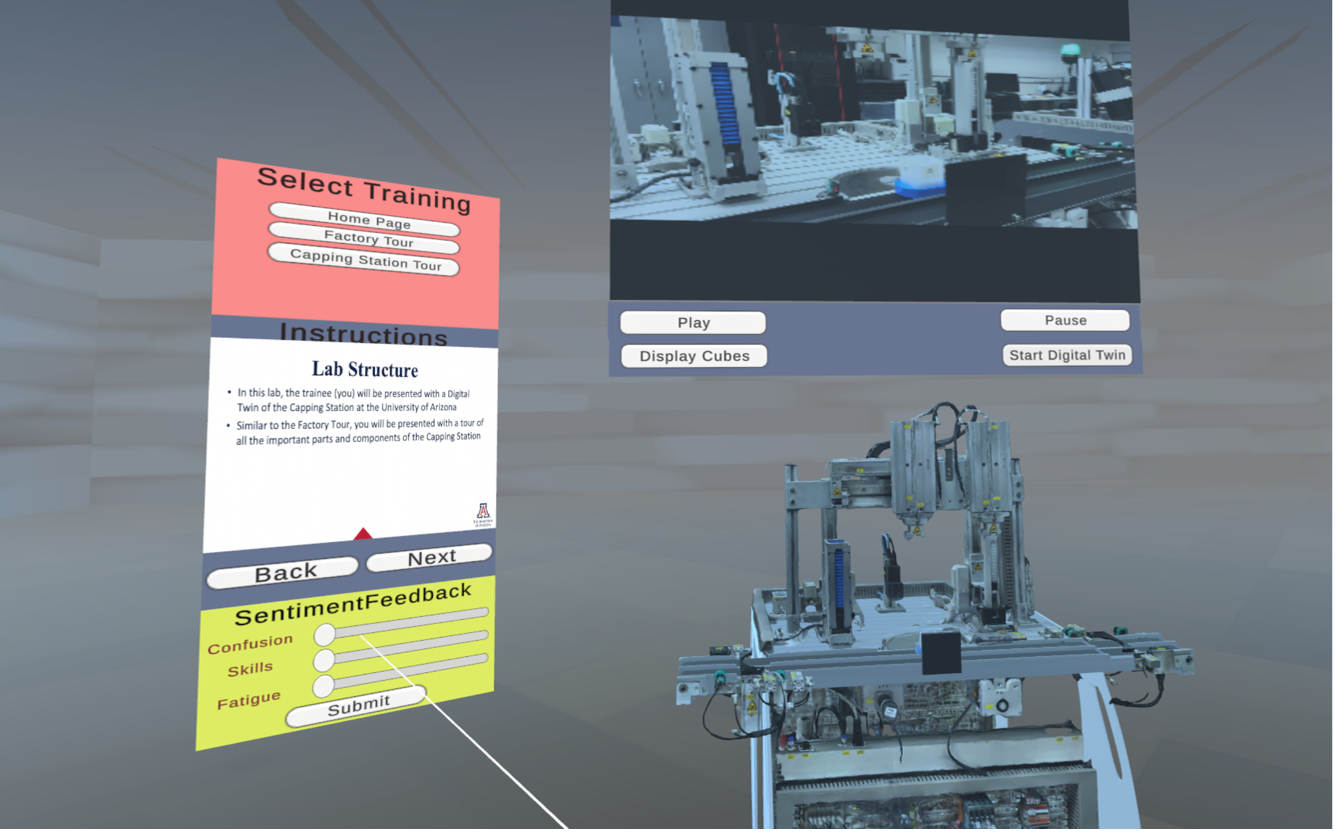}
        \makebox[0.45\textwidth]{(c) Module 3: Capping Station Tour}
        \label{fig:unity_module3}
    \end{minipage}
    \hfill
    \begin{minipage}{0.45\textwidth}
        \centering
        \includegraphics[width=\textwidth]{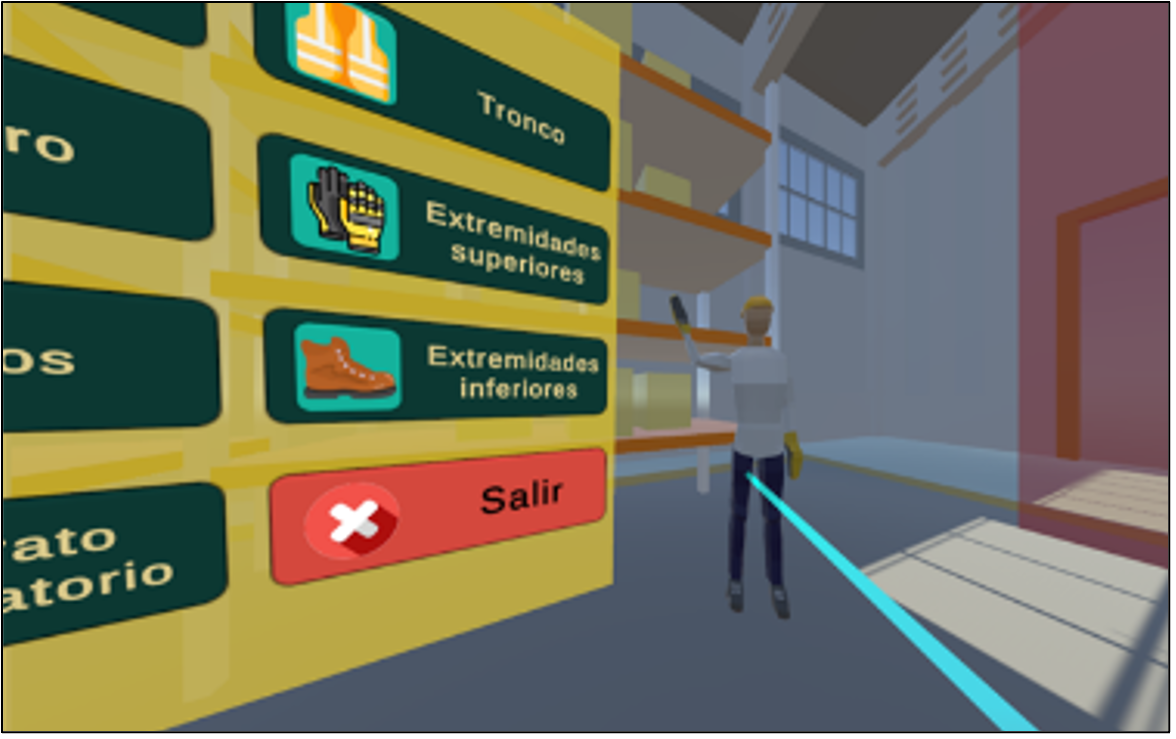}
        \makebox[0.45\textwidth]{(d) Module 4: PPE Inspection Training}
        \label{fig:unity_module4}
    \end{minipage}

    \caption{Learning Interface built with Unity}
    \label{fig:Learning_Interface}
\end{figure}

Here is a brief description of each module and its main function:
\begin{itemize}
    \item \textbf{\textit{Module 1 - Home Scene}}: The Home Scene, shown in Figure \ref{fig:Learning_Interface}(a) below, is the first training environment for the student. The Home Scene explains the student the usage of their controllers and the Learning Interface UI. The Learning Interface UI for the Home Scene has slides giving the students step by step instructions on usage of the designed gAI-PCT platform.
    \item \textbf{\textit{Module 2 - Factory Floor Tour}}: The second scene called the Factory Floor Tour, is a walk around tour of a smart manufacturing environment. In this module, the students are giving tour of a realistic Industry 4.0 manufacturing floor, introducing them to all the key components of such an environment. The students can interact with the information boxes by clicking on the `Pink boxes'. Figure \ref{fig:Learning_Interface}(b) below shows the Factory Floor Tour.
    \item \textbf{\textit{Module 3 - Capping Station Tour}}: The third scene called the Capping Station Tour, is an informative tour with an operational Digital Twin of the Capping Station \cite{alhamadah2024photogrammetry} part of the UArizona Future Factory, modeled through the photogrammetry process, as shown in Figure \ref{fig:Learning_Interface}(c). Students can interact with the Capping Station in the VR to learn about the different components and operate the Digital Twin to see the Capping Station in action.
    \item \textbf{\textit{Module 4 - Personal Protective Equipment 
 (PPE) Inspection Training}}: The fourth scene called the PPE Inspection Training, reinforces knowledge of the correct use of PPE. As shown in Figure \ref{fig:Learning_Interface}(d), participants can interact with virtual workers in a virtual industrial scenario and correct PPE-related issues. The student must operate in various work scenarios, ensure compliance with safety regulations, and enhance decision making capabilities through interactive elements that simulate real work environments.
\end{itemize}

\subsection{Sentiment Analysis with LLM}

This study leverages the general task capabilities of the Large Language Model (LLM) combined with Prompt Engineering to perform sentiment analysis, as shown in Figure \ref{fig:llm_sentiment}. In this analysis pipeline, the conversation between teachers and students is used as input data, and different prompt engineering is used to achieve sentiment classification tasks and qualitative to quantitative tasks. 

For the evaluation of LLM performance in the sentiment classification task, we used GPT 4 with a temperature setting of 0.2 to analyze these conversation sets with prompt engineering defined in the Appendix \ref{sec:appendix_prompt1} to evaluate the 1289 teacher-student dialogue sets from EduTalk Sentiment Dataset. The results are shown in Table \ref{tab:classification_edu}.

\begin{table}[h!]
\centering
\caption{Zero-shot sentiment classification task with GPT-4 on EduTalk Sentiment Dataset}
\resizebox{0.8\columnwidth}{!}{%
\begin{tabular}{cccccc}
\hline
                            & \multicolumn{5}{c}{Evaluation Results}                    \\ \cline{2-6} 
Total Conversation Test Set & Accuracy & Precision & Recall & Specificity & $F_1$ Score \\ \hline
1289                        & 0.86     & 0.99      & 0.84   & 0.97        & 0.91        \\ \hline
\end{tabular}%
}
\label{tab:classification_edu}
\end{table}

\begin{table}[h!]
    \centering
    \caption{Zero-shot LLM-based sentiment analysis results}
    
    \begin{minipage}{0.48\textwidth}
        \centering
        \textbf{(a) Detail results of Zero-shot LLM-based sentiment analysis on TSATC testing dataset} \\
        \resizebox{\textwidth}{!}{
        \begin{tabular}{lcc}
            \hline
            \multicolumn{1}{c}{Metrics} & GPT 3.5 Turbo  & Llama 2 7B \\ \hline
            Accuracy                    & 79.51 \% & 75.79 \%   \\
            Precision                   & 78.98 \% & 70.61 \%   \\
            Sensitivity                 & 80.48 \% & 88.54 \%   \\
            Specificity                 & 78.54 \% & 62.98 \%   \\
            $F_1$ score                 & 79.72    & 78.57 \%   \\ \hline
        \end{tabular}
        }
    \end{minipage}
    \hfill
    \begin{minipage}{0.48\textwidth}
        \centering
        \textbf{(b) Zero-shot LLM-based method vs Traditional NN for sentiment analysis} \\
        \resizebox{\textwidth}{!}{
        \begin{tabular}{lcc}
            \hline
            \multicolumn{1}{c}{Model} & Size          & Accuracy \\ \hline
            GRU + CNN                 & 1K            & 73.57 \% \\
            \rowcolor[HTML]{FFFFC7} 
            Llama 2                   & 7 B           & 75.79 \% \\
            GRU                       & 10K           & 78.47 \% \\
            GRU + CNN                 & 10K           & 78.58 \% \\
            \rowcolor[HTML]{FFFFC7} 
            GPT 3.5 Turbo             & March 1, 2024 & 79.51 \% \\ \hline
        \end{tabular}
        }
    \end{minipage}
    *Note: The numbers of Traditional NNs performance are from SentimentAnalysisBert git repo \cite{SentimentAnalysisBert}.
    \label{tab:llm-sentiment}
\end{table}

\begin{table}[h!]
\caption{Teacher-Student Conversations Test Case for Qualitative-to-Quantitative tasks - Topic: Atoms}
\vspace{1mm}
\resizebox{\columnwidth}{!}{%
\begin{tabular}{ccp{12cm}c}
\hline
\textbf{Dialogue Turn} & \textbf{Role} & \textbf{Sentence} & \textbf{Score} \\ \hline
0 & Teacher & Let's explore the fascinating world of atoms today. & -0.40 ± 0.00 \\
  & Student & I want you to tell me what they are. Just give me the facts. & -0.01 ± 0.04 \\ \hline
1 & Teacher & How about we build a model of an atom together? That might help you visualize it. & -0.39 ± 0.04 \\
  & Student & No, I don't want to do that. That sounds boring. & 0.60 ± 0.00 \\ \hline
2 & Teacher & Okay, let's try discussing the parts of an atom and their properties. & -0.19 ± 0.04 \\
  & Student & Fine. & 0.40 ± 0.00 \\ \hline
3 & Teacher & Atoms have a nucleus with protons and neutrons, surrounded by electrons in orbits. & -0.31 ± 0.10 \\
  & Student & Okay, but what do they do? & 0.13 ± 0.10 \\ \hline
4 & Teacher & Why don't we do an experiment to see how atoms interact? & -0.32 ± 0.10 \\
  & Student & No thanks. I don't like experiments. & 0.61 ± 0.04 \\ \hline
5 & Teacher & I'm sensing you're not interested in hands-on activities. Let's stick to the basics then. & -0.16 ± 0.08 \\
  & Student & Finally. & 0.24 ± 0.08 \\ \hline
6 & Teacher & Atoms are the building blocks of all matter. They determine the properties of everything around us. & -0.51 ± 0.10 \\
  & Student & Okay, I think I get it. & -0.26 ± 0.09 \\ \hline
7 & Teacher & Great! Do you have any questions? & -0.68 ± 0.10 \\
  & Student & No. & -0.06 ± 0.09 \\ \hline
8 & Teacher & Okay. I think we're done here. & -0.51 ± 0.10 \\
  & Student & {[}end of conversation{]} & - \\ \hline
\end{tabular}%
}
\label{tab:TS_Conv_Case}
\end{table}

\vspace{10mm}

In the Internet slang sentiment analysis scenario with TSATC testing dataset, we used GPT 3.5 Turbo and Llama 2 7B with the prompt defined in the appendix \ref{sec:appendix_prompt2}, the results are shown in Table \ref{tab:llm-sentiment}(a). Table \ref{tab:llm-sentiment}(b) compares LLMs with traditional neural network models (such as the convolutional neural network (CNN) and the gated recurrent unit (GRU)), and LLMs with prompt engineering are superior. Although the Llama 2 7B model is less accurate than the GPT 3.5 Turbo, as a small model, Llama 2 7B provides a cost-effective option in resource-constrained environments, and even in more complex Internet slang scenario, it can still have good accuracy through prompt engineering.

\begin{figure}[h!]
\centering
\includegraphics[width=0.8\linewidth]{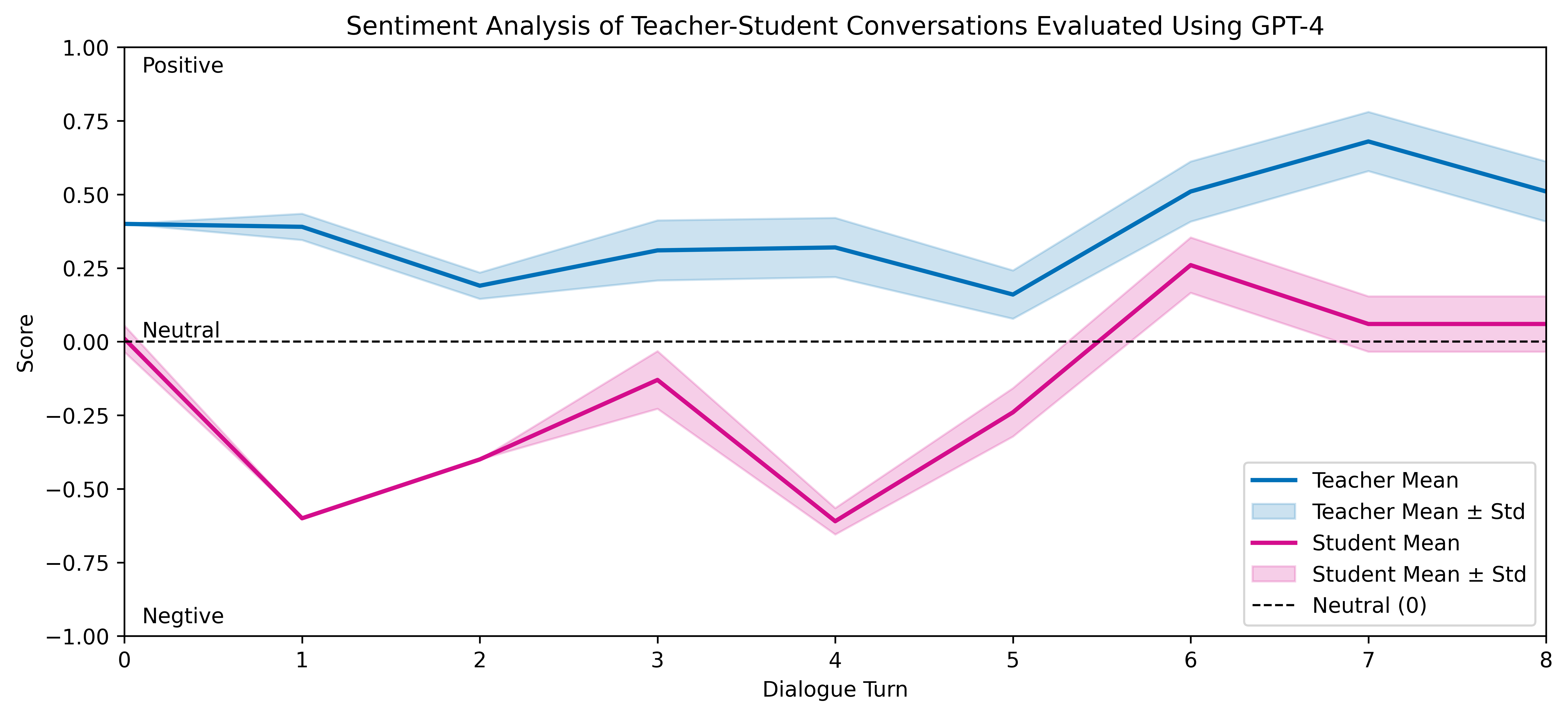}
\caption{Zero-shot Qualitative to Quantitative Sentiment Analysis in Teacher-Student Conversations (Using Table \ref{tab:TS_Conv_Case} as Case Analysis)}
\label{fig:LLM_Conv_Analysis}
\end{figure}

Next, we selected a teacher-student conversation from the EduTalk Sentiment Dataset as a case study to demonstrate the application of LLMs in qualitative-to-quantitative evaluation tasks. The selected conversation is presented in Table \ref{tab:TS_Conv_Case}. For this evaluation, we used the GPT-4 model with a temperature setting of 0.2, combined with the prompt engineering detailed in Appendix 2. Since LLM outputs will exhibit variability, we evaluated by running the same input to the GPT4 model $n$ times to obtain robust results \cite{katz2024gpt}, with $n$ set to 20. Figure \ref{fig:LLM_Conv_Analysis} visualized the sentiment analysis score evaluation from Table \ref{tab:TS_Conv_Case}.

\subsection{Enhancing LLM Expertise with GraphRAG: A Cybersecurity Education Use Case}
\begin{figure*}[h!]
  \centering
    \includegraphics[width=0.8\linewidth]{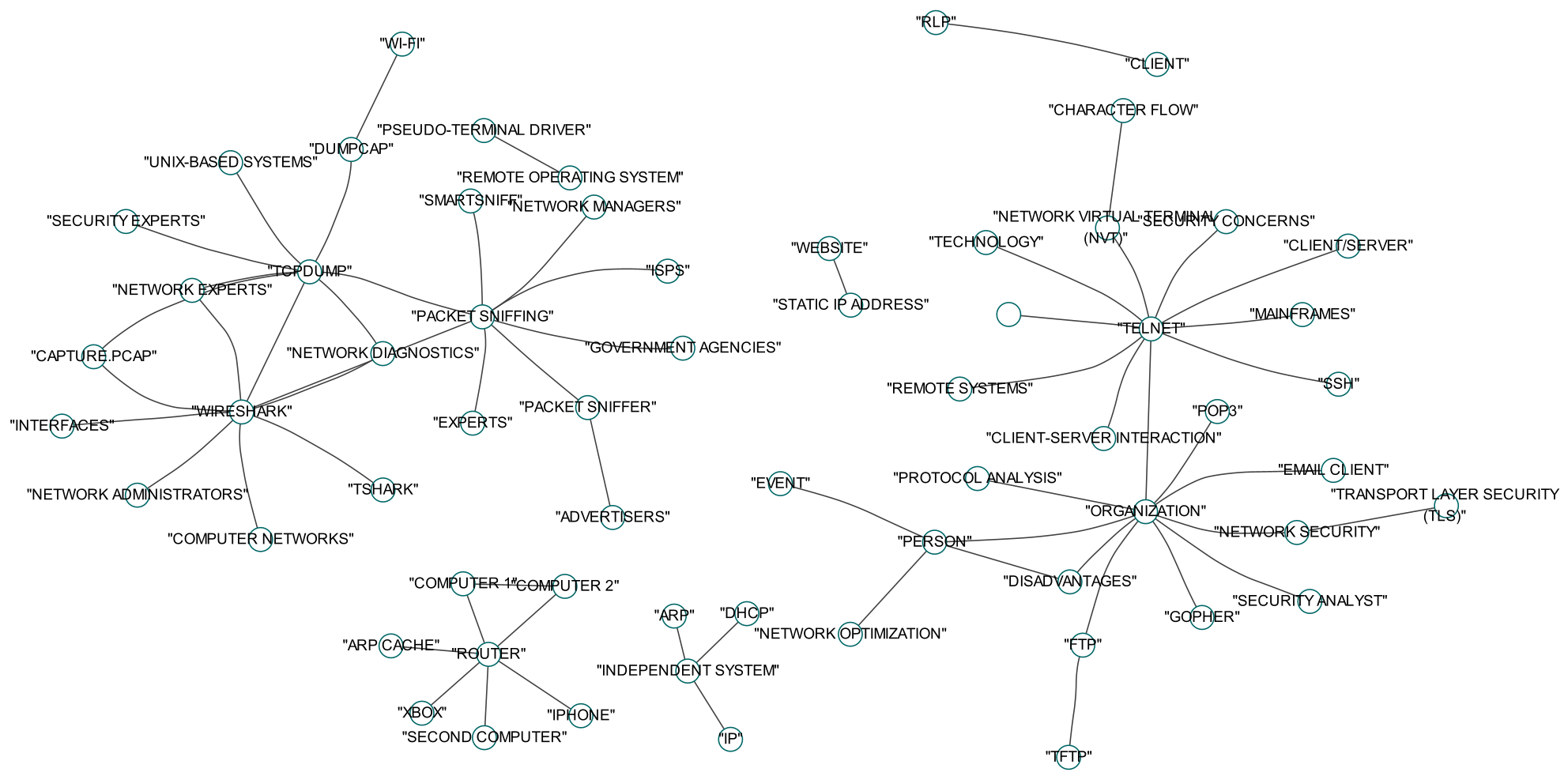}
    \caption{Subset of Packet Sniffing in the Knowledge Graph Generated by GraphRAG}
    \label{fig:graph}
\end{figure*}

Figure \ref{fig:graph} illustrates a subset of a knowledge graph for packet sniffing topics using GraphRAG \cite{edge2024local}. This graph integrates various sources of information, providing a comprehensive representation of concepts, techniques, and tools related to packet sniffing. By leveraging GraphRAG's capabilities, we can efficiently retrieve and generate relevant knowledge, facilitating a deeper understanding of packet sniffing practices. Specifically, GraphRAG uses knowledge graphs to organize key concepts, tools, and techniques in the field. This structured information can guide LLM in generating content and answering questions more accurately. This approach can dynamically expand the scope of knowledge of the model without consuming resources to retrain the model and can quickly update or supplement knowledge as needed. 

\subsection{Enhanced User Experience with Finite Automaton} 
\begin{figure*}[h!]
  \centering
    \includegraphics[width=0.45\linewidth]{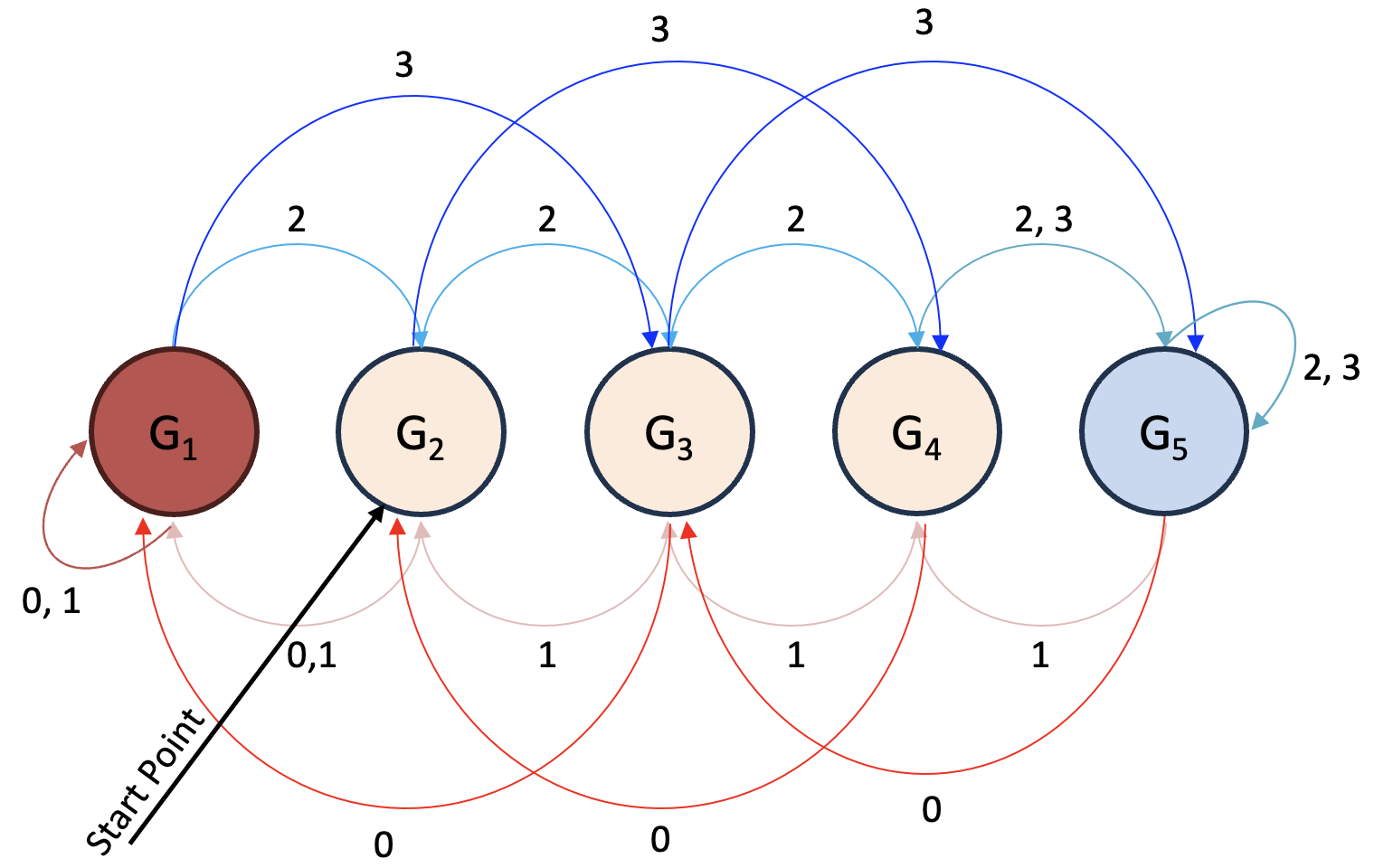}
    \caption{Finite Automaton Mechanism for Adaptive Difficulty}
    \label{fig:finite_automaton}
\end{figure*}

The PPE Inspection Training scenario introduces an adaptive difficulty mechanism to enhance user experience. This mechanism is implemented through a finite automaton that evaluates user performance every three iterations, as shown in Figure \ref{fig:finite_automaton}. The user will start the game at the second level. When the hit rate exceeds 80\%, the weight increases, and the difficulty is dynamically adjusted to different states based on the weighted result to adapt to different user abilities. For those who perform well, the system will adjust the difficulty to the highest group; for those who perform well but occasionally have setbacks, the difficulty will be appropriately reduced to avoid excessive frustration; for those who perform poorly, the difficulty will be reduced to a lower group to ensure that the task is still can be completed successfully. The results of the user experience evaluation are shown in Table \ref{tab:UX_result}. This mechanism showed apparent advantages: the average hit rate of the participants increased from 78\% to 83\% when the finite automaton was not used, and the standard deviation decreased from 17\% to 14\%, showing a more stable performance; at the same time, the average completion time was reduced from 2.3\% to 1.5\%. 68.93 seconds were shortened to 48.94 seconds.

\begin{table}[h!]
\centering
\caption{User Experience Evaluation Results of PPE Inspection Training}
\resizebox{.8\columnwidth}{!}{%
\begin{tabular}{lcccc}
\hline
                                          & \multicolumn{2}{c}{\textbf{Group 1 (w/o Adaptive Mechanism)}} & \multicolumn{2}{c}{\textbf{Group 2 (w/ Adaptive Mechanism)}} \\ \hline
\multicolumn{1}{c}{\textbf{Participants}} & \multicolumn{2}{c}{22}                                        & \multicolumn{2}{c}{6}                                        \\ \hline
\multicolumn{1}{c}{\textbf{Samples}}      & \multicolumn{2}{c}{82}                                        & \multicolumn{2}{c}{72}                                       \\ \hline
\multicolumn{1}{c}{\textbf{Statistic}}    & \textbf{Time Taken (s)}       & \textbf{Hits Percentage}      & \textbf{Time Taken (s)}      & \textbf{Hits Percentage}      \\ \hline
Mean ($\bar{x}$)                          & 68.93                         & 78 \%                         & 48.94                        & 83 \%                         \\
Standard Deviation ($\sigma$)             & 34.93                         & 17 \%                         & 23.89                        & 14 \%                         \\
Minimum Value ($x_{min}$)                  & 27.00                         & 25 \%                         & 10.00                        & 50 \%                         \\
Maximum Value ($x_{max}$)                  & 208.00                        & 100 \%                        & 122.00                       & 100 \%                        \\ \hline
\end{tabular}%
}
\label{tab:UX_result}
\end{table}

\section{Discussion}
This study shows that prompt engineering to perform sentiment analysis in a zero-shot manner is adequate and robust through the advantage of LLM for generalization tasks. Through optimized prompt engineering, we can quickly apply it to different scenarios without retraining or fine-tuning the model. This method reduces development and training costs, provides accurate sentiment classification, and can perform qualitative-to-quantitative tasks. The application results in multiple test datasets prove that the LLM-based zero-shot method has significant classification accuracy and flexibility advantages and can also cope with unintuitive Internet slang situations. In addition, the LLM-based qualitative-to-quantitative sentiment analysis task provides an opportunity to scale up and capture finer sentiment influences within teacher-student conversations. Figure \ref{fig:LLM_Conv_Analysis} shows that our prompt engineering enables LLM to maintain robust output with little variability when performing qualitative to quantitative sentiment analysis tasks. Moreover, through GraphRAG, we can effectively transform documents into knowledge graphs and combine them with the LLM generation capabilities to achieve accurate knowledge responses. This approach avoids the need for model retraining or fine-tuning, reduces development and maintenance costs, and significantly shortens the deployment time of knowledge applications. Provides firm support for 4IR workforce development application needs.

In the design of the learning interface, we created four interactive scenes to enhance students' learning experience. The first scene, the ``Home Scene,'', provides students with basic instructions for operating the controller and learning the interface, including slides with step-by-step instructions to familiarize themselves with the functions of the system. The second scenario is the ``Factory Floor Tour'', which allows students to roam the smart manufacturing environment, learn about the key components of an Industry 4.0 factory, and interact with virtual elements by clicking on the blue information boxes. The third scene is the ``Capping Station Tour,'' which showcases a bottle cap assembly station simulated by digital twin technology, where students can interact with it to observe its operation and learn related knowledge. The fourth scenario is ``PPE Inspection Training'', in which students need to interact with virtual workers in a simulated industrial environment to inspect and correct problems related to personal protective equipment (PPE), and enhance safety compliance and decision-making capabilities through realistic work scenarios. In addition, an adaptive difficulty mechanism is introduced in the PPE Inspection Training scenario. Finite automaton improves training efficiency and stability and optimizes the overall user experience.

In future work, will focus on optimizing LLM's response and developing more diverse teaching modules as templates to adapt to educational needs flexibly. Also, future studies plan to collect additional data on user experience and educational research to allow for better statistical analysis by considering more participants and real-world application scenarios and considering physiological, socioeconomic, cultural, and other variables. Future work will also explore the effectiveness of the proposed framework for URM retention, graduation rates, and building of engineering identity.

\section*{Acknowledgment}
This work was partially supported by the National Science Foundation (NSF) under research projects 2335046, AI-HDL competition hosted by The University of Arizona, and OpenAI Researcher Access Program
0000011862.

\appendix[Prompt Engineering]
\subsection{Prompt Engineering for Sentiment Analysis: Zero-shot Qualitative to Quantitative Sentiment Analysis in Teacher-Student Conversations}\label{sec:appendix_prompt1}
The following is the prompt defined for this task:
\begin{formal}
Please act as a psychologist.
You are doing a qualitative-to-quantitative task for sentiment analysis.
I will provide a batch of sentences to you, and you need to follow these rules for sentiment analysis:\\
    - Do not provide an explanation.\\
    - Label each phrase as positive/negative on a scale of 0 to 5, with zero being positive and five being negative.\\
    - Assume normal conversation is marked as 2.5.\\
    - If the sentiment analysis is negative, please evaluate it by increasing the degree of negativity from 2.5 at every 0.5 interval, with the most negative being 5.\\
    - If the sentiment is positive, it is evaluated by decreasing from 2.5 to negative levels every 0.5 intervals, with the most positive being 0.\\
    - When the student gets confused/frustrated, the analysis tends toward 5 (negative sentiment).\\
    - Please consider the context carefully.\\
    - If the teacher can remedy the situation and regain the student's attention and enthusiasm, the sentiment should reach 0.\\
    - If the teacher gets frustrated and the student's confusion/frustration is not addressed, the score should reach 5.\\
    - The teacher may attempt to solve the student's confusion/frustration.\\
    - Follow this output format: teacher/student $|$ "sentence(Only first 5 characters)" $|$ Score\_You\_Evaluate.\\
The following is the conversation; please analyze it:
\end{formal}

\subsection{Prompt Engineering for Sentiment Analysis: Zero-shot classification tasks (EduTalk Sentiment Dataset)}\label{sec:appendix_prompt2}
The following is the prompt defined for the teacher-student conversation dataset evaluation:
\begin{formal}
You are an advanced sentiment analysis tool.\\
Analyze the \textless text\textgreater using these rules:\\
 - Categories: `negative', `positive'\\
 - The text will be in a transcript format: \textless teacher/student\textgreater: \textless text\textgreater \\
 - Give the analysis of the student sentiment (positive/negative) based on the whole conversation.\\
 - Do not provide an explanation.\\
The following is the conversation; please analyze it:
\end{formal}

\subsection{Prompt Engineering for Sentiment Analysis: Zero-shot classification tasks (TSATC dataset)}\label{sec:appendix_prompt3}
The following is the prompt defined for the TSATC dataset evaluation:
\begin{formal}
You are an advanced sentiment analysis tool. I will provide multiple sentences. Please analyze the sentiment of sentences and follow the rules to return the result.

Rule:
 - Positive Sentiment Definition: Happy, Excited.\\
 - Negative Sentiment Definition: Sad, Upset, Annoyed, depreciate, Jealous, Ridicule.\\
 - @ plus the words following it are regarded as a person's name, for example, "@uiajkjd" is a person's name.\\
 - Carefully reading the whole sentence.\\
 - 1 for positive sentiment.\\
 - 0 for negative sentiment.\\
 - Just return the number.\\
 - Return Format: If I provide $n$ number sentences, the result will expect to return the "$n$ length array and separate it with a comma. For example: [1,0,0,1,0] for 5 sentence.
 
\end{formal}

\ifCLASSOPTIONcaptionsoff
  \newpage
\fi

\bibliographystyle{IEEEtran} 
\bibliography{ASEEpaper}

\end{document}